\documentclass[a4paper,fleqn]{article}
\usepackage{amsmath}
\begin{document}

\title{\textbf{On a new avatar of the sine-Gordon equation}}

\author{\textsc{Sergei Sakovich}\bigskip \\
\small Institute of Physics, National Academy of Sciences of Belarus \\
\small sergsako@gmail.com}

\date{}

\maketitle

\begin{abstract}
A chain of transformations is found which relates one new integrable case of the generalized short pulse equation of Hone, Novikov and Wang [arXiv:1612.02481] with the sine-Gordon equation.
\end{abstract}

\section{Introduction}

In this paper, we show that the nonlinear equation
\begin{equation}
u_{xt} = 1 + u - u^2 u_{xx} - u u_x^2 \label{e1}
\end{equation}
is an avatar of the sine-Gordon equation, in the sense that these two equations are related to each other by a definite chain of transformations. The nonlinear equation \eqref{e1} is interesting for the following reason.

Recently, in \cite{S1}, we studied the integrability of a class of generalized short pulse equations and found two distinct integrable cases in that class, which, up to scale transformations of variables, have the form
\begin{equation}
u_{xt} = u + \frac{1}{6} \left( u^3 \right)_{xx} \label{e2}
\end{equation}
and
\begin{equation}
u_{xt} = u + \frac{1}{2} u \left( u^2 \right)_{xx} . \label{e3}
\end{equation}
The (original) short pulse equation \eqref{e2}, which appeared first in differential geometry \cite{BRT,R} and later in nonlinear optics \cite{SW,CJSW}, is a comprehensively studied integrable equation \cite{SS1,B1,B2,SS2,SS3,M1,M2,P1,LPS1,PS,P2,FMO}. The nonlinear equation \eqref{e3}, called the single-cycle pulse equation \cite{S1} or the modified short pulse equation \cite{M3}, is a scalar reduction of the integrable two-component short pulse equation of Feng \cite{F,BS1}. These equations \eqref{e2} and \eqref{e3} are two different avatars of the sine-Gordon equation \cite{S1,SS1,SS2,SS3}. If one omits the linear term $u$ in the right-hand sides of the equations \eqref{e2} and \eqref{e3}, one gets additional two integrable equations which are avatars of linear Klein--Gordon equations \cite{S2}.

Quite recently, Hone, Novikov and Wang \cite{HNW} studied the integrability of a class of further generalized short pulse equations, with additional quadratic and cubic terms, and found, up to scale transformations of variables, seven distinct integrable cases in that class. Three of those seven cases correspond to the short pulse equation \eqref{e2}, the single-cycle pulse equation \eqref{e3}, and the celebrated integrable Ostrovsky--Vakhnenko equation (a.k.a. the reduced Ostrovsky equation, the short wave equation, the Ostrovsky--Hunter equation, or the Vakhnenko equation) \cite{O,V,VP,HW,LPS2,BS2}
\begin{equation}
u_{xt} + u u_{xx} + u_x^2 + u = 0 \label{e4}
\end{equation}
which is an avatar of the Tzitzeica equation (a.k.a. the Dodd--Bullough--Zhiber--Shabat--Mikhailov equation) \cite{MN}. The other four integrable cases found in \cite{HNW} have the form
\begin{gather}
u_{xt} = u + 2 u u_{xx} + u_x^2 , \label{e5} \\
u_{xt} = u + 4 u u_{xx} + u_x^2 , \label{e6} \\
u_{xt} = u + \left( u^2 - 4 u^2 u_x \right)_x , \label{e7} \\
u_{xt} = u + \alpha \left( 2 u u_{xx} + u_x^2 \right) + \beta \left( u^2 u_{xx} + u u_x^2 \right) , \label{e8}
\end{gather}
where $\alpha$ and $\beta$ are arbitrary nonzero constants. The equations \eqref{e5} and \eqref{e6}, as far as we know, appeared for the first time in \cite{MN}, as avatars of the sine-Gordon equation and the Tzitzeica equation, respectively. (Note that, although the equation \eqref{e5} was called the Hunter--Saxton equation in \cite{HNW}, the original Hunter--Saxton equation $u_{xt} = 2 u u_{xx} + u_x^2$ does not contain the linear term $u$, cannot be transformed to the sine-Gordon equation, and its general solution can be expressed parametrically in quadratures \cite{HS,S3}.) The equations \eqref{e7} and \eqref{e8} discovered in \cite{HNW} are completely new integrable equations, as far as we know. It has been shown in \cite{HNW} that the new equation \eqref{e7} is an avatar of the Tzitzeica equation. However, no relation between the new equation \eqref{e8} and an integrable Klein--Gordon equation has been established, and it is interesting to investigate whether there is any.

It is easy to see that the invertible transformation
\begin{gather}
u = c_1 U(X,T) + c_2 , \qquad X = c_3 x + c_4 t , \qquad T = c_5 t , \notag \\
c_1 = c_2 = - \frac{\alpha}{\beta} , \qquad c_3 = \frac{\sqrt{- \beta}}{\alpha} , \qquad c_4 = c_5 = \frac{\alpha}{\sqrt{- \beta}} \label{e9}
\end{gather}
relates the equation \eqref{e8} in the variables $u,x,t$ with the equation \eqref{e1} in the variables $U,X,T$. In the present paper, we study the new equation \eqref{e8} in its equivalent form \eqref{e1}. The variables of the equation \eqref{e1} can be considered as complex-valued ones because the variables and parameters of the equation \eqref{e8} have not gained any applied meaning yet. In Section~\ref{s2}, we show how to transform the nonlinear equation \eqref{e1} to the sine-Gordon equation. Section~\ref{s3} presents some extra results on the studied equation and its transformation. Section~\ref{s4} contains concluding remarks.

\section{Transformation} \label{s2}

The transformation
\begin{equation}
x = x(y,t) , \qquad u(x,t) = v(y,t) \label{e10}
\end{equation}
to the new independent variable $y$ brings the equation \eqref{e1} to the form
\begin{equation}
v_{yt} = \left( \frac{v_{yy}}{x_y} - \frac{v_y x_{yy}}{x_y^2} \right) \left( x_t - v^2 \right) + \frac{v_y x_{yt}}{x_y} - \frac{v v_y^2}{x_y} + ( 1 + v ) x_y , \label{e11}
\end{equation}
where $x(y,t)$ is an arbitrary function, $x_y \ne 0$. It is easy to see that the constraint
\begin{equation}
x_t = v^2 \label{e12}
\end{equation}
imposed on the function $x(y,t)$ eliminates all the second-order derivatives from the right-hand side of the equation \eqref{e11}, and we get
\begin{equation}
v_{yt} = \frac{v v_y^2}{x_y} + ( 1 + v ) x_y . \label{e13}
\end{equation}
The system of equations \eqref{e12} and \eqref{e13} is equivalent to the studied equation \eqref{e1}, in the sense that solutions of the system \eqref{e12}--\eqref{e13} determine solutions of the equation \eqref{e1} parametrically, with $y$ being the parameter, and the invariance of the system \eqref{e12}--\eqref{e13} under the transformation $y \mapsto Y(y)$ with any function $Y$ corresponds to the arbitrariness of parametrization.

Next, we introduce the new dependent variable $w(y,t)$, such that
\begin{equation}
x_y = \frac{v_y}{w} , \label{e14}
\end{equation}
which means that $w(y,t) = u_x (x,t)$. From now on, we set $v_y \ne 0$, in order to have $x_y \ne 0$ in the relation \eqref{e14}, but we keep in mind that $v_y = 0$ is admissible for the equation \eqref{e11}, where it implies $v = -1$ and corresponds to the evident constant solution $u = -1$ of the equation \eqref{e1}. The compatibility condition $(x_t)_y = (x_y)_t$ for the relations \eqref{e12} and \eqref{e14} reads
\begin{equation}
v_{yt} = \left( \frac{w_t}{w} + 2 v w \right) v_y . \label{e15}
\end{equation}
In its turn, the equation \eqref{e15} is consistent with the system \eqref{e13}--\eqref{e14} if
\begin{equation}
v = \frac{w_t - 1}{1 - w^2} . \label{e16}
\end{equation}
It is convenient to introduce the new dependent variable $z(y,t)$,
\begin{equation}
w = \tanh z . \label{e17}
\end{equation}
Then the relation \eqref{e16} takes the form
\begin{equation}
v = z_t - \cosh^2 z , \label{e18}
\end{equation}
and the equation \eqref{e15} reads
\begin{equation}
z_{ytt} - \frac{2 z_t z_{yt}}{\tanh 2z} - z_y \sinh^2 2z = 0 . \label{e19}
\end{equation}
Note that the equation \eqref{e19} is invariant under the transformation $y \mapsto Y(y)$ with any function $Y$, which means that the arbitrariness of the parameter $y$ is still not fixed.

With the help of the factor $2 z_{yt} / \sinh^2 2z$, we integrate the equation \eqref{e19} with respect to $t$ and obtain
\begin{equation}
\frac{z_{yt}^2}{\sinh^2 2z} - z_y^2 = f(y) , \label{e20}
\end{equation}
where the arbitrary function $f(y)$ appeared as a ``constant'' of integration. If $f(y) \ne 0$, we can make $f(y)$ to be any nonzero constant by means of the transformation $y \mapsto Y(y)$ with an appropriately chosen function $Y$, and this fixes the arbitrariness of $y$ up to $y \mapsto \pm y + \mathrm{constant}$. We make $f(y) = 1/4$ without loss of generality, use the new dependent variable $p(y,t)$,
\begin{equation}
z = \frac{i}{2} p , \label{e21}
\end{equation}
and the equation \eqref{e20} takes the form
\begin{equation}
p_{yt} = \pm \sqrt{1 - p_y^2} \sin p . \label{e22}
\end{equation}
Without loss of generality, we can choose the plus sign in the equation \eqref{e22}, and this fixes the arbitrariness of $y$ up to $y \mapsto y + \mathrm{constant}$. The modified sine-Gordon equation of Chen \cite{C1}
\begin{equation}
p_{yt} = \sqrt{1 - p_y^2} \sin p \label{e23}
\end{equation}
is related to the sine-Gordon equation
\begin{equation}
q_{yt} = \sin q \label{e24}
\end{equation}
by the mapping
\begin{equation}
q = p + \arcsin p_y \label{e25}
\end{equation}
which can also be expressed as the B\"{a}cklund transformation
\begin{equation}
p_y + \sin ( p - q ) = 0 , \qquad p_t - q_t + \sin p = 0 . \label{e26}
\end{equation}
Consequently, solutions of the equation \eqref{e1} are determined parametrically by the relations \eqref{e10}, where
\begin{gather}
v = \frac{i}{2} p_t - \cos^2 \frac{p}{2} , \notag \\
x_y = \left( \sqrt{1 - p_y^2} - i p_y \right) \cos^2 \frac{p}{2} , \qquad x_t = v^2 , \label{e27}
\end{gather}
$p(y,t)$ is given by the relations \eqref{e26} for any solution $q(y,t)$ of the sine-Gordon equation \eqref{e24}, and $y$ serves as the parameter.

Note, however, that not all solutions of the equation \eqref{e1} can be obtained from solutions of the sine-Gordon equation, in the way described above. Indeed, we have studied only the case of nonzero function $f(y)$ in the equation \eqref{e20} yet. When $f(y) = 0$, the equation \eqref{e20} reads
\begin{equation}
z_{yt} = \pm z_y \sinh 2z . \label{e28}
\end{equation}
The choice of plus in the equation \eqref{e28} would lead us via the relation \eqref{e18} to the excluded case of $v_y = 0$, but we have already explained above that $v_y = 0$ actually corresponds to the evident constant solution
\begin{equation}
u = -1 \label{e29}
\end{equation}
of the equation \eqref{e1}. On the contrary, the choice of minus in the equation \eqref{e28} leads to a very nontrivial solution of the equation \eqref{e1}. We integrate the minus case of the equation \eqref{e28} with respect to $y$ and obtain
\begin{equation}
z_t = - \cosh^2 z + g(t) , \label{e30}
\end{equation}
where the arbitrary function $g(t)$ appeared as a ``constant'' of integration. Then the relations
\begin{equation}
v = - 2 \cosh^2 z + g(t) \label{e31}
\end{equation}
and
\begin{equation}
x_y = - 4 z_y \cosh^2 z \label{e32}
\end{equation}
follow from the relations \eqref{e18}, \eqref{e30}, \eqref{e14} and \eqref{e17}. We integrate the relation \eqref{e32} with respect to $y$ and obtain
\begin{equation}
x = - \sinh 2z - 2z + h(t) , \label{e33}
\end{equation}
where the arbitrary function $h(t)$ appeared as a ``constant'' of integration. The relations \eqref{e33} and \eqref{e31} are consistent with the relation \eqref{e12} if $h'(t) = g(t)^2$ only. We eliminate $z(y,t)$ from the relations \eqref{e31} and \eqref{e33}, take into account the relations \eqref{e10}, and obtain in this way the corresponding solution $u(x,t)$ of the equation \eqref{e1}, in the implicit form
\begin{gather}
R + x - h + \log ( R - u + g - 1 ) = 0 , \notag \\
R = \pm \sqrt{( u - g ) ( u - g + 2 )} , \label{e34}
\end{gather}
where $g(t)$ is an arbitrary function, and $h(t) = \int g(t)^2 dt$.

Let us summarize the obtained result as follows. All solutions of the studied equation \eqref{e1}, except for the special solutions \eqref{e29} and \eqref{e34}, are determined parametrically by all solutions of the sine-Gordon equation \eqref{e24} through the relations \eqref{e10}, \eqref{e27} and \eqref{e26}.

\section{Extras} \label{s3}

In this section, we give three additional results related to the studied equation and its transformation.

The Lax pair of the equation \eqref{e1}, in the form of the over-determined first-order linear system
\begin{equation}
\Psi_x = A \Psi , \qquad \Psi_t = B \Psi , \label{e35}
\end{equation}
is determined by the matrices
\begin{gather}
A =
\begin{pmatrix}
\frac{\lambda}{2} \left( u_x^2 + 1 \right) - \lambda^3 &&
- \lambda \sqrt{1 - \lambda^2} ( u_x + \lambda ) \\[4pt]
\lambda \sqrt{1 - \lambda^2} ( u_x - \lambda ) &&
- \frac{\lambda}{2} \left( u_x^2 + 1 \right) + \lambda^3
\end{pmatrix}
, \notag \\[4pt]
B = - u^2 A +
\begin{pmatrix}
\lambda u + \frac{1}{2 \lambda} &&
\sqrt{1 - \lambda^2} u \\[4pt]
\sqrt{1 - \lambda^2} u &&
- \lambda u - \frac{1}{2 \lambda}
\end{pmatrix}
, \label{e36}
\end{gather}
where $\Psi (x,t)$ is a two-component column, and $\lambda$ is a nonzero parameter. It is easy to obtain this Lax pair by direct analysis of the compatibility condition
\begin{equation}
A_t - B_x + A B - B A = 0 \label{e37}
\end{equation}
for the system \eqref{e35}, under the assumption that $A ( u_x )$ and $B ( u , u_x )$ are traceless $2 \times 2$ matrices. The Lax pair \eqref{e35}--\eqref{e36} of the equation \eqref{e1} can be transformed to the second-order scalar Lax pair given in \cite{HNW} for the equation \eqref{e8}, by means of the transformation \eqref{e9} and a gauge transformation of $\Psi$.

The system \eqref{e12}--\eqref{e13}, written in the polynomial form as
\begin{equation}
x_y v_{yt} - v v_y^2 - ( 1 + v ) x_y^2 = 0 , \qquad x_t - v^2 = 0 , \label{e38}
\end{equation}
does not pass the Painlev\'{e} test for integrability of nonlinear partial differential equations \cite{WTC,T}, for the reason of non-dominant logarithmic branching of solutions. To show this, we use the expansions
\begin{gather}
v = v_0 (y) \phi^\sigma + \dotsb + v_n (y) \phi^{\sigma + n} + \dotsb , \notag \\
x = x_0 (y) \phi^\tau + \dotsb + x_n (y) \phi^{\tau + n} + \dotsb \label{e39}
\end{gather}
to represent solutions of the system \eqref{e38} near a non-characteristic hypersurface $\phi (y,t) = 0$, with $\phi = t + \psi (y)$ and $\psi' (y) \ne 0$. We find that the dominant singular behavior of solutions is determined by
\begin{equation}
\sigma = \tau = -1 , \qquad v_0 = \pm 1 , \qquad x_0 = -1 , \label{e40}
\end{equation}
and that the positions of resonances, where arbitrary functions of $y$ can enter the expansions \eqref{e39}, are
\begin{equation}
n = -1 , 1 , 2 , \label{e41}
\end{equation}
with $n = -1$ corresponding to the arbitrariness of $\psi (y)$. At the resonance $n = 1$, we find that $v_1 = 0$, the function $x_1 (y)$ is not determined, and the nontrivial compatibility condition $\psi' (y) = 0$ appears which contradicts to the initial assumption that the hypersurface $\phi (y,t) = 0$ is non-characteristic. As the result, solutions of the system \eqref{e38} cannot be represented by purely Laurent-type expansions \eqref{e39}--\eqref{e40}, the actual singular expansions of solutions must contain some non-dominant logarithmic terms, starting from the terms proportional to $\phi^0 \log \phi$, and the integrable system \eqref{e38} does not pass the Painlev\'{e} test for integrability. It is easy to show, however, that the actual singular expansions of solutions of the system \eqref{e38} are
\begin{gather}
v = \pm \phi^{-1} - \frac{1}{2} + v_2 (y) \phi + \dotsb , \notag \\
x = - \phi^{-1} \mp \log \phi + x_1 (y) + \left( \frac{1}{4} \pm 2 v_2 (y) \right) \phi + \dotsb , \label{e42}
\end{gather}
where $x_1 (y)$, $v_2 (y)$ and $\psi (y)$ (in $\phi$) are arbitrary functions. Note that there is only one logarithmic term in the expansions \eqref{e42}, namely, the one already shown in $x$, therefore the singular expansions of $v$, $x_y$ and $x_t$ contain no logarithmic terms at all. Some integrable nonlinear equations with dominant logarithmic branching of solutions are known in the literature \cite{C2,P3}. The system \eqref{e38} is the first example of an integrable nonlinear equation with non-dominant logarithmic branching of solutions, as far as we know.

The nonlinear equation \eqref{e28} can be completely solved in quadratures. (The choice of plus or minus in the equation does not matter, up to $t \mapsto - t$.) Note that we did not solve the minus case of the equation \eqref{e28},
\begin{equation}
z_{yt} = - z_y \sinh 2z , \label{e43}
\end{equation}
when we obtained the implicit special solution \eqref{e34} of the studied equation \eqref{e1}. Instead, we used the possibility to eliminate $z$ from the expressions for $v$ \eqref{e31} and $x$ \eqref{e33}, without any knowledge of what is the function $z(y,t)$ exactly. It is easy to see, however, that the function $r(y,t)$ given by the relation
\begin{equation}
\exp r = ( \exp (-2z) )_y \label{e44}
\end{equation}
satisfies the Liouville equation
\begin{equation}
r_{yt} = \exp r \label{e45}
\end{equation}
with the well-known general solution
\begin{equation}
r = \log \frac{2 a_y b_t}{( a + b )^2} , \label{e46}
\end{equation}
where $a(y)$ and $b(t)$ are arbitrary functions. We use the expression \eqref{e46} for $r$, integrate the relation \eqref{e44} with respect to $y$, and obtain
\begin{equation}
\exp (-2z) = \frac{- 2 b_t}{a + b} + c(t) , \label{e47}
\end{equation}
where the arbitrary function $c(t)$ appeared as a ``constant'' of integration. The function $z (y,t)$ determined by the relation \eqref{e47} satisfies the equation \eqref{e43} if the functions $b(t)$ and $c(t)$ satisfy the relation
\begin{equation}
\frac{b_{tt}}{b_t} = \frac{c_t}{c} + \frac{c}{2} - \frac{1}{2 c} . \label{e48}
\end{equation}
As the result, the general solution of the equation \eqref{e43} has the form
\begin{equation}
z = - \frac{1}{2} \log \left( \frac{- 2 b_t}{a + b} + c \right) , \label{e49}
\end{equation}
where the functions $a(y)$ and $c(t)$ are arbitrary, and
\begin{equation}
b(t) = \int{ c \, \exp \left( \frac{1}{2} \int{\left( c - \frac{1}{c} \right) dt} \right) dt} . \label{e50}
\end{equation}
Let us note, however, that we were unable to locate the completely solvable equation \eqref{e43} in the classifications \cite{ZS,KPZ} which were most relevant to the subject in our opinion.

\section{Conclusion} \label{s4}

In this paper, we studied one new integrable equation of Hone, Novikov and Wang \cite{HNW}, in its equivalent form \eqref{e1}. We found a chain of transformations which relates the integrable equation \eqref{e1} with the sine-Gordon equation. The result means that all the seven integrable cases of the generalized short pulse equation studied in \cite{HNW} are avatars of two integrable nonlinear Klein--Gordon equations, namely, the sine-Gordon equation and the Tzitzeica equation. The chain of transformations we obtained for the equation \eqref{e1} is rather complicated one, it leads to the sine-Gordon equation via the Chen's modified sine-Gordon equation, and in this respect the equation \eqref{e1} is similar to the so-called sine-Rabelo equation \cite{SS3}. On the other hand, the equation \eqref{e1} is very similar in its form to the single-cycle pulse equation \eqref{e3} (the only essential difference being the constant term) whose envelope soliton can be as short as only one cycle of its carrier frequency. Therefore we believe that the new integrable equation of Hone, Novikov and Wang, in its original form \eqref{e8}, or in the form \eqref{e1}, or in a different convenient form, can gain a definite applied meaning in problems of nonlinear wave propagation in media with cubic nonlinearities.

\end{document}